\documentclass[prb,aps,twocolumn,showpacs]{revtex4}%
\usepackage{amsmath}
\usepackage{amsfonts}
\usepackage{amssymb}
\usepackage{graphicx} 
\usepackage{bm}

\pdfoutput=1

\begin{document}

\title{Effect of doping on irreversible Campbell length in MgB$_2$ single crystals }

\author{C. Martin}
\affiliation{Ames Laboratory and Department of Physics and Astronomy, Iowa State University, Ames, IA 50011}
\author{M. D. Vannette}
\affiliation{Ames Laboratory and Department of Physics and Astronomy, Iowa State University, Ames, IA 50011}
\author{R. T. Gordon}
\affiliation{Ames Laboratory and Department of Physics and Astronomy, Iowa State University, Ames, IA 50011}
\author{R. Prozorov}
\email[Corresponding author: ]{prozorov@ameslab.gov}
\affiliation{Ames Laboratory and Department of Physics and Astronomy, Iowa State University, Ames, IA 50011}

\author{J. Karpinski}
\affiliation{Laboratory for Solid State Physics, ETH, 8093 Zurich, Switzerland}
\author{N. D. Zhigadlo}
\affiliation{Laboratory for Solid State Physics, ETH, 8093 Zurich, Switzerland}

\date{18 July 2008}

\begin{abstract}
We have measured small-amplitude rf penetration depth $\lambda(H,T)$ in pure and
C, Li and (Li+C) doped single crystals of MgB$_2$. The effect of doping on the
critical temperature T$_c$ and on the upper critical field H$_{c2}$ was found
to be in good agreement with previous results. We report the presence of
clear signatures of irreversibility in $\lambda(H,T)$, associated with the peak effect.
Carbon doping enhances the signature and shifts its position on the H-T phase diagram
to higher temperatures. Contrary to this, Li substitution has the effect of suppressing
the peak effect to lower temperatures. Analysis of both zero field (ZFC) and field cooled (FC) measurements suggests that the hysteresis associated with the peak effect is due
to macroscopic screening supercurrents, $j$, generated during the magnetic field ramp.
\end{abstract}


\maketitle

\section{\label{sec:Intro}Introduction}
MgB$_2$ is a two-band (two quasi 2D, $\sigma$-bands and two 3D, $\pi$-bands), two-gap ($\sigma$-gap, $\Delta_{\sigma}\approx$ 78 K and $\pi$-gap, $\Delta_{\pi}\approx$ 29 K) superconductor with high critical temperature T$_{c}\approx$ 39 K for an s-wave superconductor~\cite{Nagamatsu01}. These characteristics have made MgB$_2$ very attractive for advancing theoretical understanding of superconductivity in two-gap systems.  Theory has shown that the two gap nature has a significant effect on the H$_{c2}$(T) phase diagram, which is governed by the ratio between the intraband diffusivities $D_{\sigma}/D_{\pi}$~\cite{Gurevich03}.  The high transition temperature is advantageous for practical applications, however the potential is limited by the relatively low values for the upper critical fields, particularly for the c-axis orientation, H$_{c2}^{\| c}$ = 3-6 T. 

Atomic substitution is a method to affect different physical quantities like carrier concentration, density of states, lattice parameters, interband and/or intraband scattering rates in a controlled way.  Examining how these quantities affect the superconducting properties is important both for theoretical understanding and for improving the potential for practical use of
MgB$_2$.  Moreover, doping may influence significantly the vortex lattice structure and dynamics. Successful substitutions with Al~\cite{Slusky01,Karpinski2005}, Mn~\cite{Xu01,Rogacki2006} and Li~\cite{Zhao01} for Mg, and C~\cite{Takenobu01,Kazakov2005} for B, as well as co-dopings~\cite{Xu03,Karpinski07} have been reported.

Hall probe measurements of local ac-magnetic susceptibility revealed the existence of the \textit{peak effect (PE)}, i.e. a maximum of  the diamagnetic screening in a certain region of the H-T phase diagram, in single crystals of both pure and C-doped and Al-doped MgB$_{2}$~\cite{Pissas02, Pissas04, Pissas07}. This effect had been previously observed in both high-T$_c$ and low-T$_c$ superconductors, but theoretically its origin is not yet fully understood and several possible explanations have been put forth~\cite{Mikitik01}. Larkin and Ovchinnikov proposed that the effect is caused by a softening of the elastic moduli of the flux-line lattice (FLL) in the vicinity of H$_{c2}$~\cite{Larkin79}. Alternatively, it was suggested that as the elastic moduli softens, the pinning forces overcome the elastic ones, marking a transition from weak to strong pinning~\cite{Blatter04}. 

A completely different interpretation was generated by the strong experimental evidence for the peak effect in disordered systems, like Nb~\cite{Ling01, Park03}. Earlier theoretical work on vortex matter in superconductors with disorder induced pinning had claimed that instead of the Abrikosov lattice, i.e. long range ordered vortex state, a quasi-ordered, \textit{Bragg glass} phase stabilizes~\cite{Giamarchi95}. Therefore the peak effect feature was interpreted as a transition from Bragg-glass to a disordered vortex phase. The hysteresis of ac-susceptibility which develops at the PE, particularly the low diamagnetic screening state which results after zero field cooling (ZFC) the sample and applying magnetic field, was regarded as evidence for the Bragg glass vortex phase in MgB$_2$~\cite{Pissas02, Pissas04}. A recent experimental study~\cite{Prozorov03} using the same technique involved in the present work found similar hysteresic behavior of the rf penetration depth in a series of high and low temperature superconductors. However, Ref.~[\onlinecite{Prozorov03}] suggests that ramping the magnetic field after ZFC gives rise to macroscopic screening supercurrents, $j$, which shift the vortices into a state of inhomogeneous distribution, in accordance with the critical state (Bean) model. The resulting state is a displaced vortex lattice that is thought to disappear when field cooling (FC) the sample due to a relaxation of screening currents, therefore giving rise to hysteresis.   

The present work studies the effect of C, Li and (Li+C) doping on the vortex dynamics in single crystals of MgB$_2$. The results of our study are two-fold: (1) We show that the same anomaly of rf susceptibility described in the previous paragraph is present in both pure and doped MgB$_2$ and it coincides on the H-T phase diagram with the measured peak effect from local ac-magnetic susceptibility from Hall probe experiments~\cite{Pissas02, Pissas04}.  (2) We show that the C doping at B site enhances the maximum of the rf-diamagnetic screening by increasing the number of pinning centers. Contrary to this, when Li is substituted for Mg, the peak effect is suppressed to lower temperatures, suggesting a slight reduction of pinning.  Further, studies with both FC and ZFC are shown to be in good agreement with the proposed scenario of relaxing supercurrents after field ramp. The model from Ref.~[\onlinecite{Prozorov03}] is discussed in the context of collective pinning theory as a possible explanation for both the shape of the anomaly and the history dependence associated with the peak effect.

\section{Experimental Details}
The samples used for the present study were single crystals of pristine MgB$_2$,  12$\%$ Li substituted
at Mg site (Mg$_{0.88}$Li$_{0.12}$B$_2$), 4.8$\%$ and 7.5$\%$ C substituted at B site (MgB$_{1.904}$C$_{0.096}$ and MgB$_{1.85}$C$_{0.15}$ respectively), and 12$\%$ Li and  6$\%$ C  co-doped 
(Mg$_{0.88}$Li$_{0.12}$B$_{1.88}$C$_{0.12}$). The crystals were grown using a high pressure cubic anvil technique, described in Refs.~[\onlinecite{Karpinski03, Karpinski07}]. The current samples come from the same batches as those studied in Ref.~[\onlinecite{Karpinski07}] where detailed information about the crystal structure and stoichiometry are provided.

The rf susceptibility was measured by placing the sample inside the coil of a self-resonant LC circuit,
powered by a tunnel diode. The resonant frequency of the Tunnel Diode Resonator (TDR),
$2\pi f_{0}=1/\sqrt{LC}$ is about 14 MHz. The experimental data were obtained using a $^3$He refrigerator with a 90 kOe superconducting magnet. For the present work, both the dc and the rf magnetic fields were applied along the crystallographic $c$-axis, so that the penetration depth was measured in the $ab$-plane ($\lambda_{ab}$).

When a superconducting material is placed inside the inductor it produces a change in resonant frequency proportional to the susceptibility, and in turn, to the penetration depth of the sample, $\Delta f\propto\Delta\lambda$. Given that the sample shapes were almost rectangular, following the procedure described
in Ref.~[\onlinecite{Prozorov00}], the rf susceptibility is obtained from:
\begin{equation}
\Delta f(T)\propto4\pi\chi=\frac{1}{1-N}\left[1-\frac{\lambda}{R}\tanh\left(\frac{R}{\lambda}\right)\right],
\label{equ: Susc}
\end{equation}
where $N$ is the demagnetization factor and $R$ is the effective dimension of the sample as described by Ref.~[\onlinecite{Prozorov00}].

At low temperatures and for applied magnetic fields H$\gg$ H$_{c1}$, the penetration depth, and thus the dynamic susceptibility, is $\lambda^{2}=\lambda^{2}_{L}+\lambda^{2}_{v}$, where $\lambda_{L}$ is the London penetration depth and
$\lambda_{v}$ is the vortex contribution. In the pinning regime, the main contribution to the total penetration depth is given by the vortex motion, which in turn reduces to the Campbell pinning penetration depth $\lambda_C$, resulting from the oscillating Lorentz force exerted by the rf screening currents on the vortices:
\begin{equation}
\lambda_{C}^{2}=\frac{\phi_{0}B}{4\pi\kappa_{p}},
\label{equ: Campbell}
\end{equation}
where $\phi_{0}$ is the flux quantum and $\kappa_{p}$ is the Labusch parameter, which measures the curvature of the pinning potential, $V(r)$, $\kappa_{p}=\vert{d^{2}V(r)/dr^{2}\vert}$.

It should be mentioned that, besides technical details, there are two major differences between the TDR and other ac techniques, such as the mutual inductance or based on Hall sensor. One is the frequency range; the mutual inductance technique uses low frequencies, up to kHz, whereas the TDR is used in our case at MHz values. However, this difference should not be important as long as both frequencies are significantly lower than the pinning frequency, most commonly in the GHz range. A more important difference is the magnitude of the excitation ac magnetic field. As it can be seen from Eq.~\ref{equ: Campbell}, by shaking the vortices with a probe ac magnetic field, i.e. ac screening current, the curvature of the pinning potential at the vortex location can be determined. The excursion that the vortices make while oscillating depends on the magnitude of the ac magnetic field. In our experimental set-up, the applied rf field was estimated to be 5 mOe $\leq b_{rf}\leq$ 10 mOe. For comparison, the ac magnetic fields used in two studies of the peak effect in MgB$_2$~\cite{Pissas02, Pissas04} are larger than 1 Oe, which is at least 100 times larger than in the current work. The implication is that the amplitude of the vortex oscillation around the equilibrium position are at least 100 times larger in Refs.~[\onlinecite{Pissas02, Pissas04}], thus the curvature of the pinning potential is not probed locally, but rather is averaged over the length of the vortex excursion.

\section{Results and Discussion}
\begin{figure}[tb]
\begin{center}
\includegraphics[keepaspectratio=1,width=9cm]{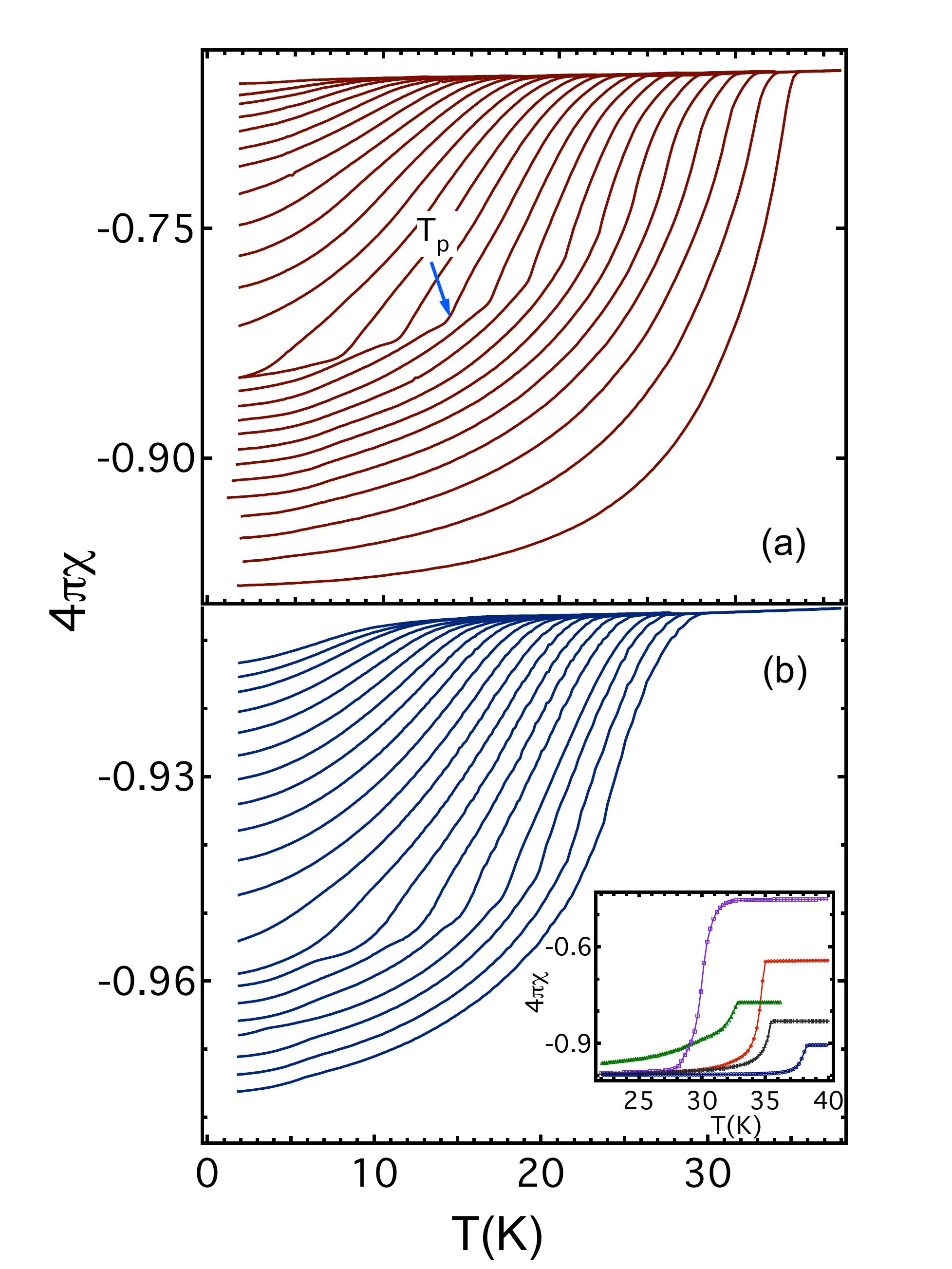}
\end{center}
\caption[ Susceptibility Undoped]
{\linespread{1}\normalfont (a) The rf susceptibility $\chi(T)$ of Mg$_{0.88}$Li$_{0.12}$B$_2$ in applied magnetic fields from 2 kOe to 54 kOe in steps of 2 kOe. The position of the peak effect is marked as  T$_p$.  (b) The rf susceptibility $\chi(T)$ of the undoped MgB$_2$ in magnetic field from 14 kOe to 52 kOe in 2 kOe steps. Inset: Zero field $\chi(T)$ showing the entire transition for the studied samples: MgB$_2$ (circles, T$_{c}\approx$38.5K), MgB$_{1.904}$C$_{0.096}$ (crosses, T$_{c}\approx$35.5K), Mg$_{0.88}$Li$_{0.12}$B$_2$ (diamonds, T$_{c}\approx$35K), MgB$_{1.85}$C$_{0.15}$ (triangles, T$_{c}\approx$32.8K),  and Mg$_{0.88}$Li$_{0.12}$B$_{1.88}$C$_{0.12}$ (squares, T$_{c}\approx$32K).} 
\label{fig: Susc_Und}
\end{figure}

\begin{figure}[tb]
\begin{center}
\includegraphics[keepaspectratio=1,width=9cm]{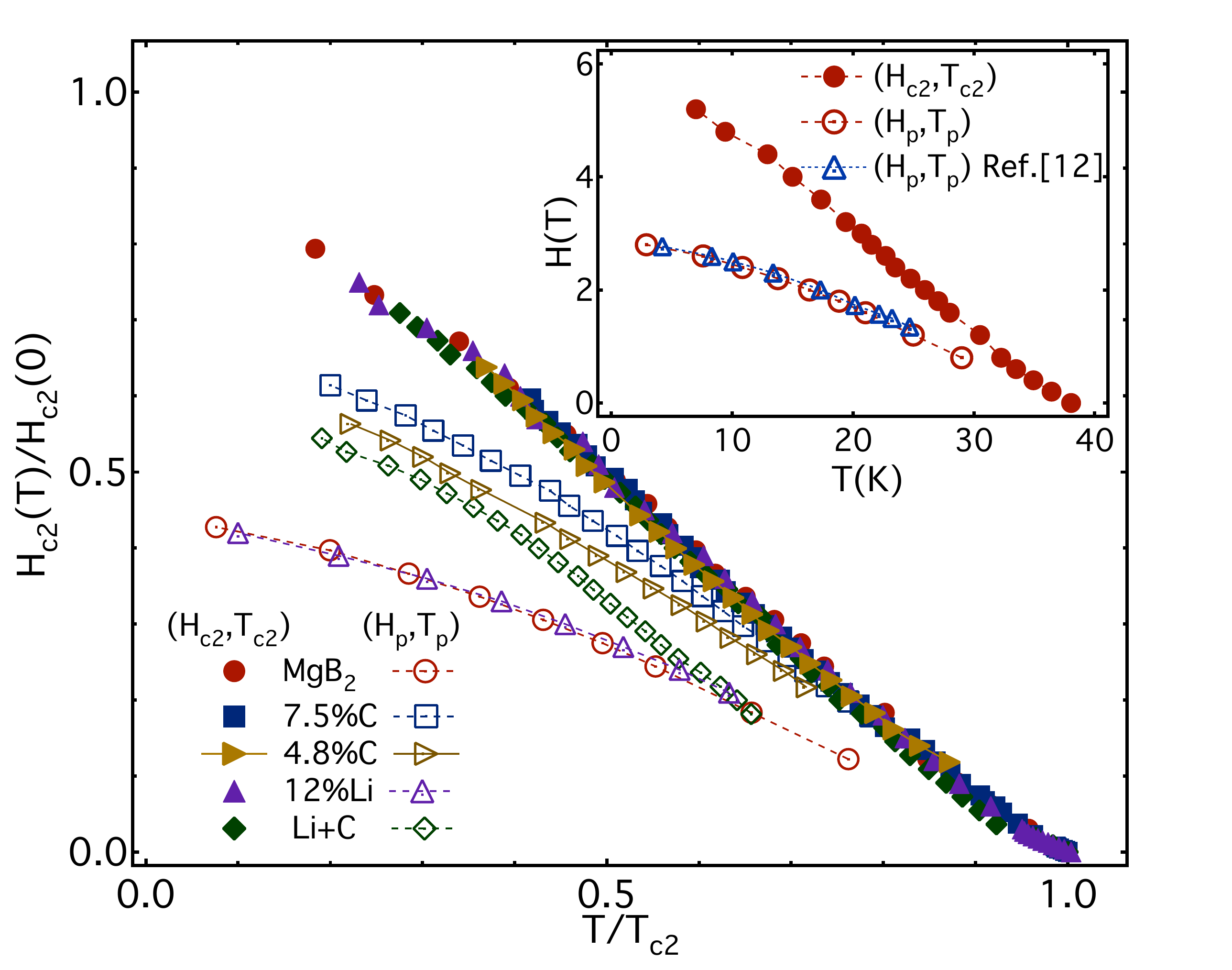}
\end{center}
\caption[The reduced H$_{c2}$(T) phase diagram]
{\linespread{1}\normalfont The reduced phase diagram for the studied samples, showing the values for 
H$_{c2}$(T) and the peak effect H$_{p}$(T), normalized to H$_{c2}$(0) and to T$_c$. The symbols are explained in the legend. Inset: The absolute values of H$_{c2}$(T) and  H$_{p}$(T) for the undoped
MgB$_{2}$. The values for the peak effect H$_{p}$(T) from Ref.~[\onlinecite{Pissas02}] are displayed for comparison.} 
\label{fig: HT}
\end{figure}

Figure~\ref{fig: Susc_Und} shows the dynamic magnetic susceptibility for 12$\%$ Li substituted (a) and undoped (b) MgB$_{2}$. It can be seen on both figures that for applied magnetic fields
$\mu_{0}H\geq$ 12 kOe, there is an anomaly, i.e. a small kink in $\chi(T)$. For the pristine
MgB$_{2}$, the position of the kink,  $\left(H{_p}, T{_p}\right)$, on the H-T phase diagram is in good agreement with the observed peak effect from Ref.~[\onlinecite{Pissas02}] (see inset of Fig.~\ref{fig: HT}). When 12$\%$ of Li is substituted for Mg, the critical temperature decreases to T$_{c}\approx$35K, as previously reported in Ref.~[\onlinecite{Karpinski07}]. However, it is evident from Fig.~\ref{fig: Susc_Und}(a) that the size of the anomaly is not significantly affected by doping. The normalized H-T phase diagram in Fig.~\ref{fig: HT} shows that the PE is suppressed to lower temperatures in the Li-doped crystal. A possible explanation is that doping with Li may actually reduce disorder by filling Mg vacancies possibly present in pure MgB$_2$ crystals, which are believed to be  magnesium  deficient.  As a consequence, the pinning of vortices in the Mg layers is reduced, and lower energy thermal fluctuations overcome the pinning forces, smearing the PE. This is supported by crystal structure results, where a narrowing of some x-ray reflection peaks was found in lithium doped samples~\cite{Karpinski07}. Also, the upper critical field H$_{c2}$(0) is slightly lower in Li doped crystals indicating less disorder.  Further de Haas -van Alphen measurements to determine the mean free path would help with clarifying this hypothesis. 

\begin{figure}[tb]
\begin{center}
\includegraphics[keepaspectratio=1,width=9cm]{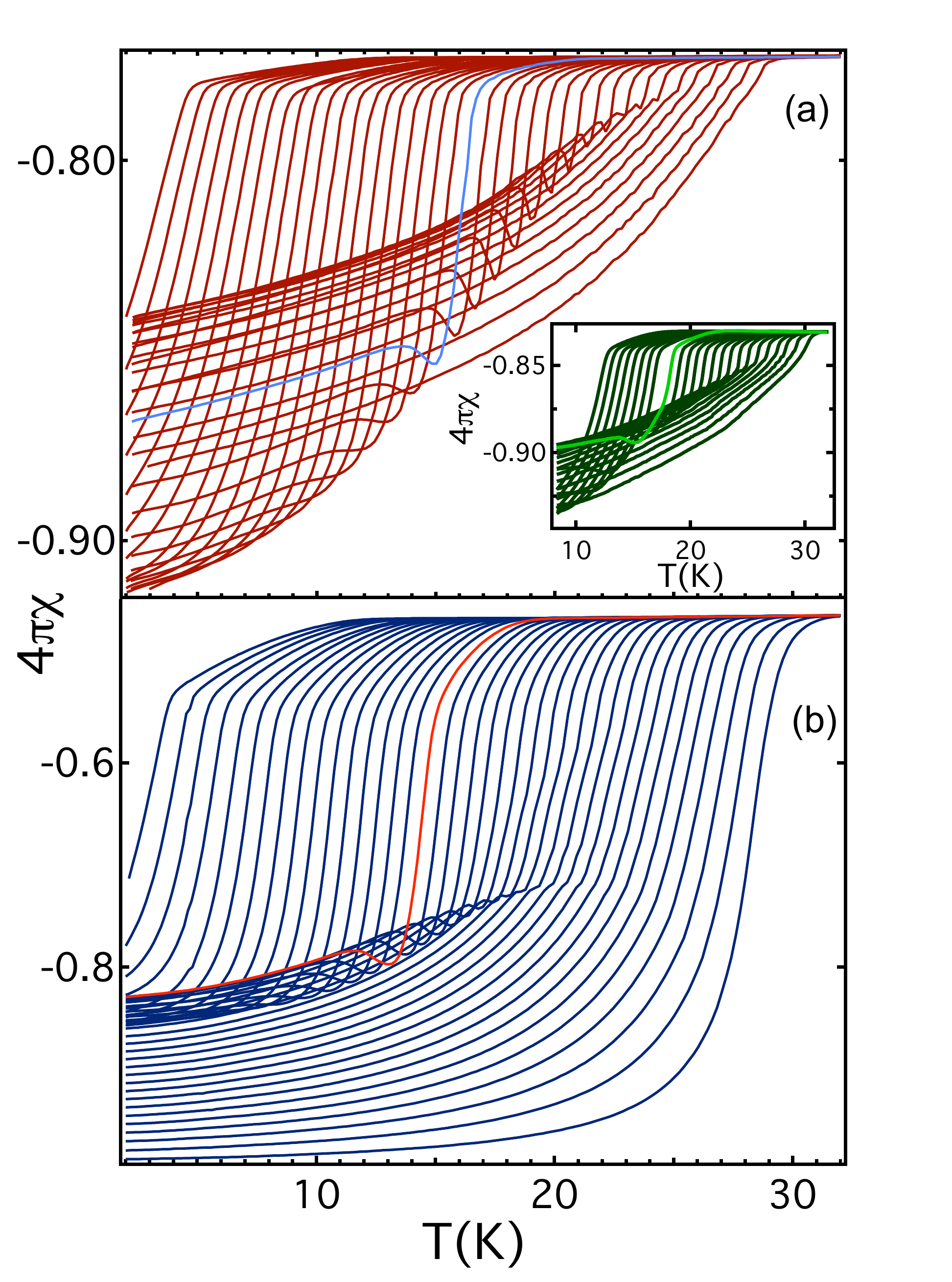}
\end{center}
\caption[ Susceptibility MgB$_2$ doped 12$\%$Li]
{\linespread{1}\normalfont (a) $\chi(T)$ of  MgB$_{1.85}$C$_{0.15}$ at magnetic fields from 10 kOe to 86 kOe in 2 kOe steps. Inset: $\chi(T)$ of  MgB$_{1.904}$C$_{0.096}$ at magnetic fields from 10 kOe to 58 kOe in 2 kOe steps. (b) $\chi(T)$ for the co-doped sample, Mg$_{0.88}$Li$_{0.12}$B$_{1.88}$C$_{0.12}$ in fields between 2 and 78 kOe, in steps of 2 kOe, showing very similar behavior like the C doping. The traces at 46 kOe (and at 40 kOe for the inset) have been highlighted for more clarity of the peak effect.} 
\label{fig: Susc_C}
\end{figure}

Contrary to the Li doping, when C is substituted for B, the effect on the vortex dynamics is significantly stronger, as shown in Fig.~\ref{fig: Susc_C}(a). A sharper, more pronounced maximum of the diamagnetic screening develops, similar to previous observations from local ac susceptibility at lower frequency~\cite{Pissas04}. We now compare two different C concentrations: 4.8$\%$ (inset of Fig.~\ref{fig: Susc_C}(a)) and 7.5$\%$ (Fig.~\ref{fig: Susc_C}(a)). First, we note that at approximately the same field where the PE appears, there are two transitions at $T_c$ visible in both samples: a lower one associated with the onset of superconductivity $T_{c2}$, and an upper transition temperature which we define as $T_{c3}$, quite possibly due to surface superconductivity. The evolution of $T_{c2}$ with applied magnetic field agrees well with that from resistivity measurements~\cite{Masui04}. Figure~\ref{fig: HT} suggests that the position of the peak effect shifts to higher magnetic fields as the C content increases and it appears at significantly larger fields than in the undoped or Li-doped MgB$_2$. 

\begin{figure}[tb]
\begin{center}
\includegraphics[keepaspectratio=1,width=9cm]{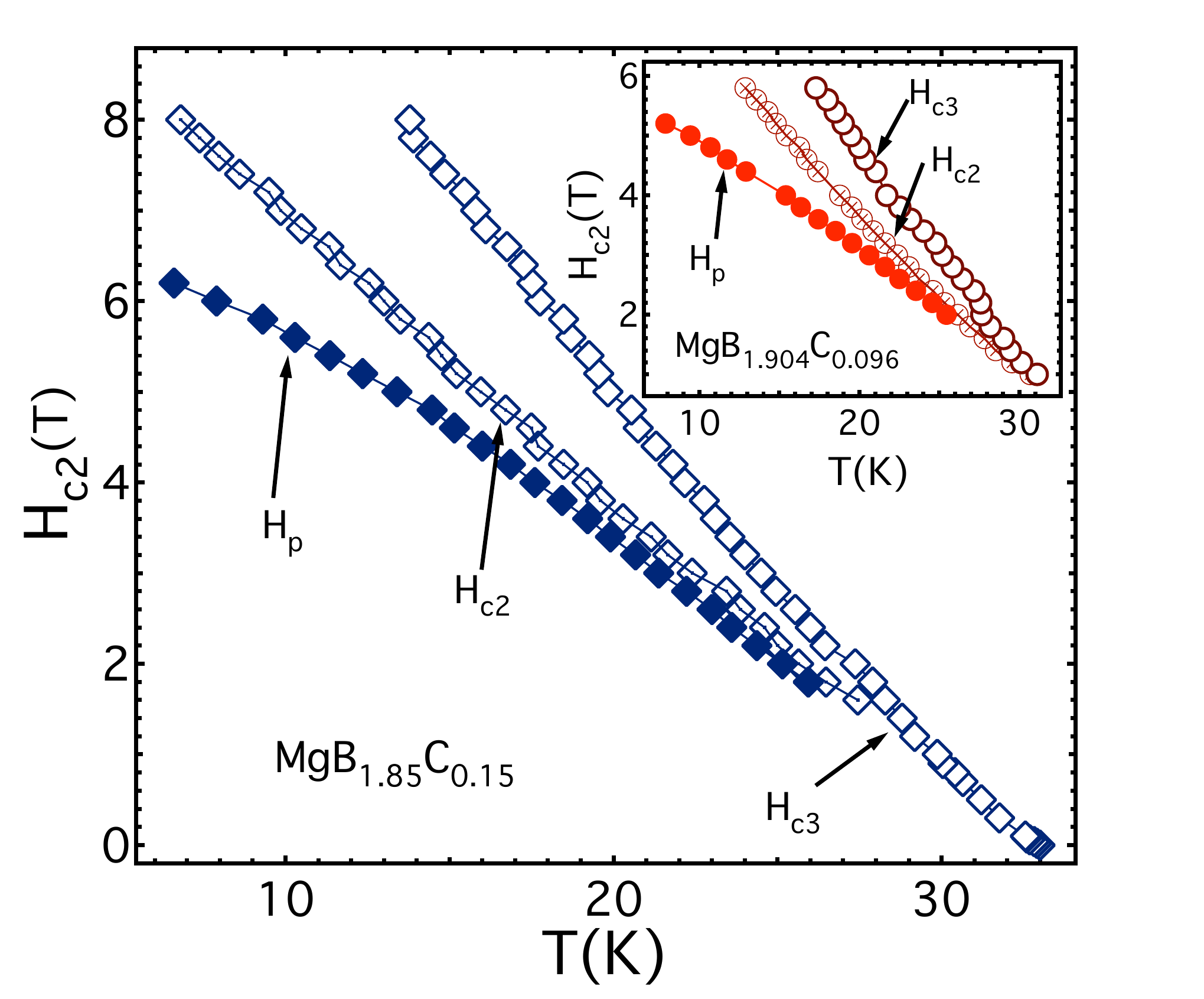}
\end{center}
\caption[H$_{c2}$(T) phase diagram for C doped]
{\linespread{1}\normalfont The phase diagram for the C-doped samples: MgB$_{1.85}$C$_{0.15}$ (main graph) and  MgB$_{1.904}$C$_{0.096}$ (inset). The three main curves representing H$_{c2}$(T), H$_{c3}$(T) and H$_{p}$(T) are shown in both graphs.} 
\label{fig: HT_C}
\end{figure}

For clarity, we have plotted separately, in Fig.~\ref{fig: HT_C}, the phase diagram for the two C-doped samples. It is important to notice that the three main curves, $H_{p}(T)$, $H_{c2}(T)$ and $H_{c3}(T)$ do not appear to meet at a tricritical point in our data both for 4.8$\%$
and 7.5$\%$ C doping. This is in contrast with the results obtained in Nb single crystals~\cite{Park03}, despite the strong similarity between the magnetic ac-susceptibility of MgB$_2$ and Nb. Our results support the interpretation that the peak effect is suppressed by thermal fluctuations~\cite{Lee08} and not necessarily at a tricritical point where a possible Bragg glass disordering transition occurs. Moreover, closer examination of our data in Fig.~\ref{fig: HT_C}, reveals that the PE becomes more pronounced and survives to higher temperatures in the sample with the higher concentration of C.  We attribute this to the increase in pinning strength as carbon content increases.

The susceptibility of the co-doped sample, Mg$_{0.88}$Li$_{0.12}$B$_{1.88}$C$_{0.12}$, is consistent with the cumulative effect of Li and C (Fig.~\ref{fig: Susc_C}(b)). The peak effect is marked by a pronounced dip of $\chi(T)$ and there are two transitions visible at T$_c$, similar to what is seen in the carbon doped samples (Fig.~\ref{fig: Susc_C}(b)).  Furthermore, the presence of Li shifts the position of the peak effect to lower temperatures on the reduced phase diagram in Fig.~\ref{fig: HT}. The co-doping strongly reduces T$_c$ (inset of Fig.~\ref{fig: Susc_Und}(b)).  A detailed analysis and possible explanation for the evolution of T$_c$ with doping is given in Ref.~[\cite{Karpinski07}].

\begin{figure}[tb]
\begin{center}
\includegraphics[keepaspectratio=1,width=9cm]{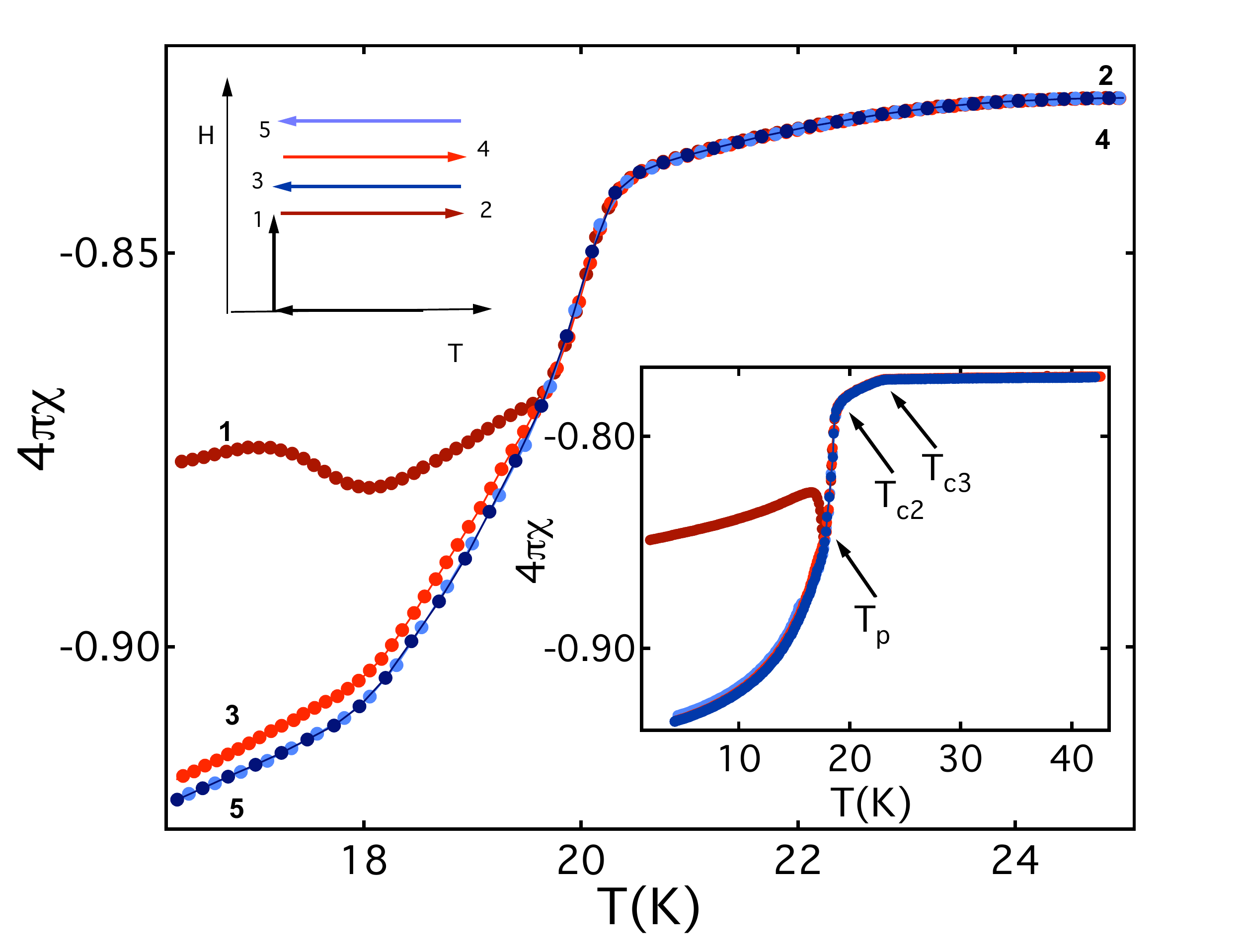}
\end{center}
\caption[Hysteresis at the peak effect]
{\linespread{1}\normalfont History dependence of $\chi(T)$ in MgB$_{1.904}$C$_{0.096}$ (main figure)
and MgB$_{1.85}$C$_{0.15}$ (lower inset) for applied magnetic field H=35 kOe, following a sequence explained in the text and sketched in the upper inset.
} 
\label{fig: Hyst}
\end{figure}

An important characteristic of the peak effect is the presence of hysteresis in dc magnetization and ac susceptibility. In order to verify and understand the nature of the hysteresis, we performed measurements of rf susceptibility both by zero field cooling-field warming (ZFC-FW) and field cooling-field warming (FC-FW) the samples, following a scenario schematically illustrated in the upper inset of Fig.~\ref{fig: Hyst}. The results for both C-doped crystals are shown in
Fig.~\ref{fig: Hyst}. After ZFC the sample to base temperature $T\ll T_{c}$, the magnetic field was ramped to a certain value resulting in state 1. Then, as the temperature is increased on curve 1 to 2, the susceptibility shows a diamagnetic peak at T$_p$, before the transition to the normal state at T$_{c2}$ or T$_{c3}$. Upon FC the sample from 2 to 3, the strong peak is no longer present, but at the same temperature
T$_p$, the susceptibility develops strong hysteresis, resulting in a stronger diamagnetic screening at base temperature. By further increasing the temperature from 3 to 4, the susceptibility follows almost the same path as 2$\rightarrow$3, showing only a weak \textit{kink} at T$_{p}$. Cooling back in field, the trace 4$\rightarrow$5 overlaps with  2$\rightarrow$3. The strong maximum in rf diamagnetic response and the strong hysteresis at  T$_{p}$ are only present during the first temperature sweep after ramping the magnetic field, after which it only manifests as a kink in $\chi(T)$. Therefore, we can conclude that the state resulting after ZFC to base temperature and increasing magnetic field has a much weaker diamagnetic susceptibility then the subsequent warming and cooling the sample in applied magnetic field (FC). 

\begin{figure}[tb]
\begin{center}
\includegraphics[keepaspectratio=1,width=8cm]{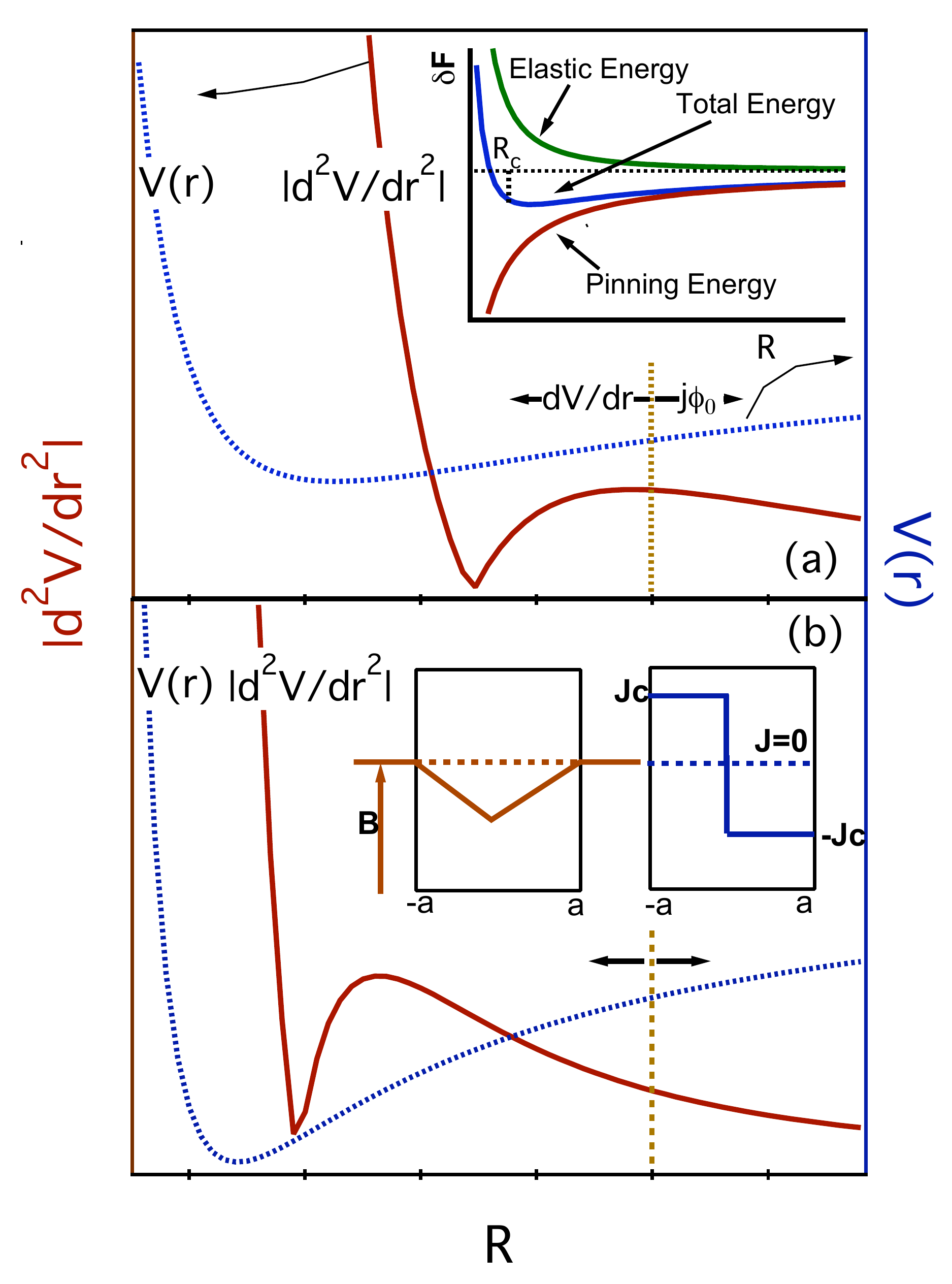}
\end{center}
\caption[Qualitative picture of the pinning potential]
{\linespread{1}\normalfont The pinning potential V(r) and the second derivative $\vert{d^{2}V(r)/dr^{2}\vert}$ (Labusch parameter) of a FLL for arbitrary fixed values of C$_{66}$, C$_{44}$, $\xi$, f  in Eq.~\ref{equ: PinEn}, and two different 
values for n: n$_1$ (a) and 3n$_1$(b). In both graphs it is sketched  the additional displacement due to the Lorentz force produced by screening supercurrents $j$. Inset (a): The total free energy of a vortex bundle as a function of distortion length and the contribution from elastic (first two terms of Eq.~\ref{equ: PinEn}) and pinning forces (last term of Eq.~\ref{equ: PinEn}). Inset (b): Field and current distribution within the Bean model when the field is ramped after ZFC (continuous lines) and after FC (dashed lines).}
\label{fig: Pin}
\end{figure}

Although, the rf Campbell penetration depth cannot be used to rule out the existence of the Bragg glass in MgB$_2$, we propose as a possible explanation for the hysteresis associated with the peak effect a model based on the idea suggested in Ref.~[\onlinecite{Prozorov03}] for an individual 
pinning potential, and extend it to the collective pinning regime. Within the collective pinning theory~\cite{Larkin79}, following notations from Ref.~[\onlinecite{Tinkham96}], for a density of pinning centers, $n$, the free energy density of the vortex bundle (FLL) within the correlation volume, $V_c$,
can be written as the sum of the pinning and the elastic energy:
\begin{equation}
\delta F=\frac{1}{2}C_{66}\left(\frac{\xi}{R_{c}}\right)^{2}+\frac{1}{2}C_{44}\left(\frac{\xi}{L_{c}}\right)^{2}-
f\xi\frac{n^{1/2}}{V_{c}^{1/2}},
\label{equ: PinEn}
\end{equation}
where $C_{66}$ is the shear modulus and $C_{44}$ is the tilt modulus of the FLL, f is the pinning force, $\xi$-the correlation length, and $R_c$ and $L_c$ describe average distances over which the FLL is distorted due to shear and tilt, respectively. As detailed knowledge of pinning in MgB$_2$ is not yet available, the inset of Fig.~\ref{fig: Pin}(a) shows a qualitative plot for arbitrary values of the constants in Eq.~\ref{equ: PinEn}. During the ramping of the magnetic field after ZFC the sample, within the Bean model, there will be a macroscopic distribution $B(x)$ of magnetic field inside the sample, accompanied by a current distribution $\mu_{0}j=dB(x)/dx$ (inset of Fig.~\ref{fig: Pin}(b)). These currents exert a Lorentz force ($f_{L}=j\phi_{0}$) on vortices causing additional bending of the vortex bundle from its equilibrium (minimum energy) configuration. The displacement takes place until the Lorentz force is compensated by the pinning forces ($dV/dr=j\phi_{0}$). The situation is sketched in Fig.~\ref{fig: Pin}(a) and (b). At the peak effect, whether because the pinning becomes strong enough or because the elastic modulus softens, the vortices relax to equilibrium. The Campbell penetration depth will then map the curvature of the pinning potential during the relaxation process. As seen in Fig.~\ref{fig: Pin}(a), the curvature can change sign, so that the absolute value shows a small kink, very similar to that observed in the penetration depth of the undoped or Li-doped MgB$_2$. If in Eq.~\ref{equ: PinEn}, the number of pinning sites, $n$ is increased (Fig.~\ref{fig: Pin}(b)) while all the other parameters are kept the same, the kink becomes more pronounced, and the behavior is very similar to that observed in our C-doped MgB$_2$ data. Therefore, the model suggests that substitution with C at B sites, adds disorder in B planes, increasing the number of pinning
centers. On the other hand, the fact that there is no clear change in the shape of the anomaly for Li substitution is consistent with
the idea that Li does not increase the density of pinning sites. Indeed, it slightly lowers the pinning site density as was inferred from Fig.~\ref{fig: HT}.   

When the sample is further FC, the field distribution inside is uniform (dashed lines of the inset in Fig.~\ref{fig: Pin}(b)), the supercurrent $j=0$ and a stable FLL is established. This explains both the absence of a clear dip in rf susceptibility and the absence of hysteresis for subsequent warming and cooling under applied magnetic field. Then, the Campbell penetration depth features only a kink at the peak effect, consistent with a transition from a weak (strong) to strong (weak) pinning regime as the sample is warmed (cooled) in field.

It must be noted that our findings are similar to those of Refs.~[\onlinecite{Pissas02, Pissas04}], but in previous reports there is still a visible minimum in ac susceptibility after FC the sample. However, both in Ref.~[\onlinecite{Pissas02}] and Ref.~[\onlinecite{Pissas04}], the ac magnetic field is at least two orders of magnitude larger than in our experimental setup. As explained in the experimental section, our technique probes the curvature of the pinning potential much more locally (over a few $\AA$ based on an estimate similar to that in Ref.~[\onlinecite{Prozorov03}]) whereas in other techniques the amplitude of the vortex oscillation is much larger.

\section{Conclusions}
In conclusion, we have studied the small-amplitude rf susceptibility of pure, C, Li and C-Li doped MgB$_2$ single crystals. At the peak effect, the rf susceptibility $\chi(T)$ shows a kink, similar to that observed in local ac susceptibility measurements. Doping with C was found to have a strong influence on the peak effect, consistent with increasing the density of pinning centers, whereas Li substitution seems to slightly reduce the pinning in MgB$_2$. This could have practical implications in finding the proper way to achieve high critical currents and critical magnetic fields in MgB$_2$.
We propose that the history dependence associated with the peak effect is not necessarily a signature of a transition to a vortex glass phase.  Rather it may be the result of additional distortion of the FLL caused by screening currents induced during a magnetic field ramp. Such a model may have broader implications, possibly explaining the history effects observed in ac susceptibility measurements on other superconductors.

\begin{acknowledgments}
We would like to thank V. G. Kogan for very stimulating discussions and R. Khasanov for help with the samples. Work at the Ames Laboratory was supported by the Department of Energy-Basic Energy Sciences under Contract No. DE-AC02-07CH11358. Work at ETH was supported by the Swiss National Science Foundation through NCCR pool MaNEP. R. P. acknowledges support from Alfred P. Sloan Foundation.
\end{acknowledgments}

\end{document}